\documentclass[superscriptaddress,twocolumn]{revtex4}

\pdfoutput=1 

\usepackage{graphicx,epstopdf,float}
\usepackage{bbm, bm, mathbbol, amsmath, longtable, amssymb,bbold}
\usepackage{color}
\usepackage{braket}
\newcommand{\beq}{\begin{equation}}
\newcommand{\eeq}{\end{equation}}

\definecolor{DarkGreen}{rgb}{0,.5,0}

\begin{document}

\title{Interfacing GHz-bandwidth heralded single photons with a room-temperature Raman quantum memory}

\author{P.~S.~Michelberger}
\affiliation{Clarendon Laboratory, University of Oxford, Parks Road, Oxford OX1 3PU, UK}

\author{T.~F.~M.~Champion}
\affiliation{Clarendon Laboratory, University of Oxford, Parks Road, Oxford OX1 3PU, UK}

\author{M.~R.~Sprague}
\affiliation{Clarendon Laboratory, University of Oxford, Parks Road, Oxford OX1 3PU, UK}

\author{K.~T.~Kaczmarek}
\affiliation{Clarendon Laboratory, University of Oxford, Parks Road, Oxford OX1 3PU, UK}

\author{M.~Barbieri}
\affiliation{Clarendon Laboratory, University of Oxford, Parks Road, Oxford OX1 3PU, UK}

\author{X.~M.~Jin}
\affiliation{Clarendon Laboratory, University of Oxford, Parks Road, Oxford OX1 3PU, UK}
\affiliation{Department of Physics and Astronomy, Shanghai Jiao Tong University, Shanghai 200240, PR China}

\author{D.~G.~England}
\affiliation{Clarendon Laboratory, University of Oxford, Parks Road, Oxford OX1 3PU, UK}
\affiliation{National Research Council of Canada, Ottawa, Ontario K1A 0R6, Canada}

\author{W.~S.~Kolthammer}
\affiliation{Clarendon Laboratory, University of Oxford, Parks Road, Oxford OX1 3PU, UK}

\author{D.~J.~Saunders}
\affiliation{Clarendon Laboratory, University of Oxford, Parks Road, Oxford OX1 3PU, UK}

\author{J.~Nunn}
\affiliation{Clarendon Laboratory, University of Oxford, Parks Road, Oxford OX1 3PU, UK}

\author{I.~A.~Walmsley}
\affiliation{Clarendon Laboratory, University of Oxford, Parks Road, Oxford OX1 3PU, UK}

\maketitle

\textbf{Photonics is a promising platform for quantum technologies \cite{OBrien:2009aa}. However, photon sources \cite{Eisaman2011} and two-photon gates \cite{OBrien:2003aa} currently only operate probabilistically. Large-scale photonic processing will therefore be impossible without a multiplexing strategy to actively select successful events.
High time-bandwidth-product quantum memories --- devices that store and retrieve single photons on-demand --- provide an efficient remedy via active synchronisation \cite{Sanguard2011, Nunn:2013ab}. Here we interface a GHz-bandwidth heralded single-photon source and a room-temperature Raman memory with a time-bandwidth product exceeding 1000.
We store heralded single photons and observe a clear influence of the input photon statistics on the retrieved light which agrees with our theoretical model. The preservation of the stored field's statistics is limited by four-wave-mixing noise, which we identify as the key remaining challenge in the development of practical memories for scalable photonic information processing.}

Temporal multiplexing with quantum memories works by storing the successful outputs of heralded, probabilistic photonic primitives, such as sources and gates. 
On-demand retrieval can then be used to actively synchronise multiple operations, thereby making repeat-until-success strategies scalable \cite{Sanguard2011, Nunn:2013ab}. A key figure of merit for this task is the time-bandwidth product $B = \tau \delta$, where $\tau$ is the memory lifetime and $\delta$ is the memory acceptance bandwidth. 
$B$ corresponds to the maximum number of possible processor steps that can occur within the memory lifetime. Dramatic enhancements in the scaling of photonic networks become possible with $B \sim 1000$ \cite{Nunn:2013ab}. The very long storage times \cite{Longdell:2005aa,Dudin:2013aa,Heinze:2013aa} and large time-bandwidth products \cite{Reim2011,Bao:2012aa} obtainable by optical storage in atoms make these systems prime candidates for temporal synchronisation devices.

The first step towards memory-based synchronisation of single-photon generation is the storage of single photons. This has been demonstrated by spin-wave storage in cold atomic ensembles \cite{Bimbard:2014aa,Choi:2008aa,Zhang:2011aa} and by pre-programmed atomic-frequency-comb echoes in cryogenic rare-earth-doped crystals \cite{Rielander:2014aa,Bussieres:2014aa}. 
Room-temperature storage \cite{Jensen:2010aa,Eisaman:2005aa} and retrieval \cite{Appel:2008aa} of non-classical light has been achieved with narrow bandwidths on the order of a few MHz. However, a quantum memory suitable for large-scale temporal multiplexing is still elusive. Key desiderata are broad acceptance bandwidths, on-demand storage and retrieval, and room-temperature operation with low noise.

In this paper we report the storage and retrieval of GHz-bandwidth heralded photons, generated by spontaneous parametric down-conversation (SPDC), in a room-temperature caesium-vapour Raman memory, for which $B \gtrsim 1000$ \cite{Reim2010, Reim2011}. Feed-forward control of the memory operation, which is crucial for on-demand synchronisation tasks, is implemented by triggering the memory on the detection of a herald photon. The performance of our system is characterised by measuring the autocorrelation of the fields transmitted through, and retrieved from the memory. We present a simple theoretical model and identify four-wave mixing as the sole significant noise source; suppressing this process will make it possible to use this technology to construct large-scale photonic networks.

\begin{figure*}
\begin{center}
\includegraphics[width=.8\textwidth]{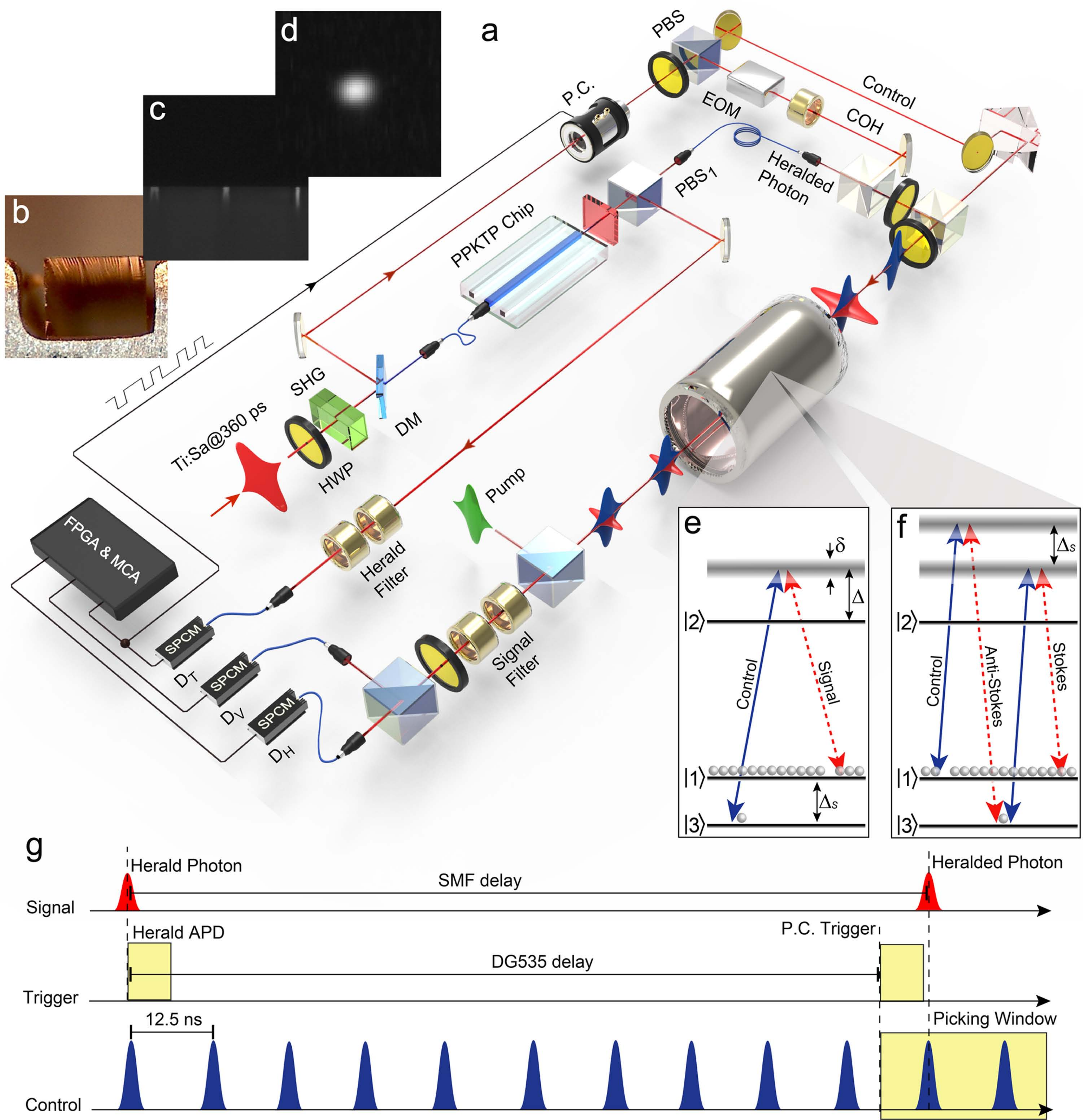}
\caption{\textbf{Schematic of the experiment}. 
\textbf{a} Experimental setup (Methods, Supplementary Information):
HWP: half-wave plate, SHG: second-harmonic generation, DM: dichroic mirror, PBS: polarising beam splitter, PC: Pockels cell, EOM: electro-optic modulator, COH: coherent-state input signal, SPCM: single-photon counting module. \textbf{b}: PPKTP waveguide chip input face. \textbf{c}: waveguides of 4-2~$\mu$m width (left to right). \textbf{d}: EM-CCD image showing the spatial mode of the SPDC photons from the 3~$\mu$m waveguide used in the experiment.
\textbf{e} Raman memory scheme (see Methods): A signal pulse, in two-photon resonance with a strong control and detuned by $\Delta$ from the excited state in Cs, is stored in an ensemble of initially spin-polarised atoms (state $\ket{1}$). 
During storage the signal is read in by exciting a spin-wave coherence \cite{Fleischhauer2000, Reim2010} between the initial state $\ket{1}$ and the storage state $\ket{3}$ in the Cs. Read-out, by reapplication of the control, transfers the atoms back to the initial state $\ket{1}$ and releases the signal.
\textbf{f} FWM noise process (main text and Supplementary Information):
The control field can couple to the ground state $\ket{1}$, driving spontaneous anti-Stokes scattering, which produces spurious atomic excitations that can be retrieved as noise.
\textbf{g} Feed-forward timing diagram (see Methods): Herald detection triggers the Pockels cell to pick a pair of control pulses from the Ti:Sa output. Appropriate time delays ensure control and signal pulses enter the memory simultaneously.
}
\label{fig:setupV61}
\end{center}
\end{figure*}

\begin{figure}
\begin{center}
\includegraphics[width=\columnwidth]{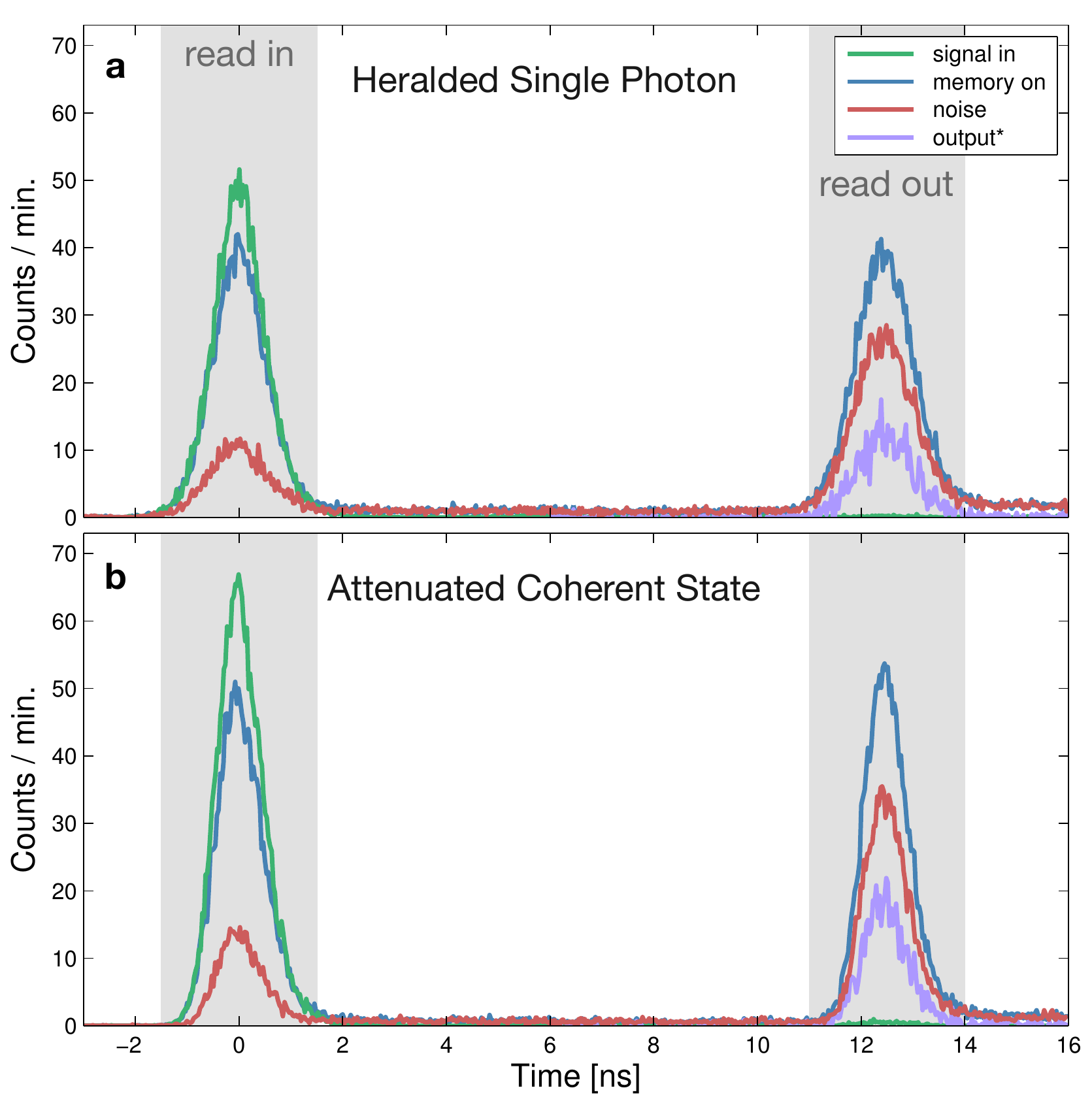}
\caption{\textbf {Temporal intensity profiles demonstrating memory.} Storage for $t=12.5 \, \text{ns}$ of \textbf{a} single-photon input 
signals, and \textbf{b} weak coherent states, with approximately equal input photon number per pulse of $N_\text{in}= \eta_\mathrm{herald}=0.22$ and $N_\text{in}^\text{COH} = 0.23$, respectively. Green traces show the input signal field transmitted through the memory when the control field is blocked, i.e. the memory is off (signal only). Red lines show the noise emitted by the memory in the read-in and read-out time bin, when the input signal is blocked (control only).
Blue lines display the pulses in each time bin when the memory is on, i.e. control field and input signal are applied together (signal + control).
This includes signal, transmitted through or recalled from the memory, along with any added noise. Subtracting the noise from this trace in the read-out bin gives the effective retrieved signal (lilac). This is used in direct comparison with the input signal (signal only) to determine the memory efficiency (Supplementary Information).}
\label{fig:Store_retrieve_traces}
\end{center}
\end{figure}

Our memory is based on transient 
Raman absorption in a Cs vapour cell at $70^{\circ}\text{C}$ \cite{Nunn2007,Reim2010,Reim2011, England2012,Sprague:2014aa} (Fig. \ref{fig:Store_retrieve_traces} and Methods). 
The memory's large acceptance bandwidth of $\delta = 1.2$~GHz, set by the control pulse, allows the memory to be directly interfaced with a traveling-wave spontaneous parametric down-conversion (SPDC) source without the need to enhance the spectral brightness by an optical cavity \cite{Fekete:2013}. 
Our source is based on type-II SPDC in a periodically-poled potassium-titanyl-phosphate (PPKTP) waveguide (Fig. \ref{fig:setupV61} \textbf{a}). 
The single-pass waveguide is pumped by a pulse of approximately 260 ps duration at a wavelength of 426 nm. 
SPDC produces photons in pairs, thus we herald the presence of a signal photon to be stored in the memory by detecting its partner idler photon ($\text{D}_\text{T}$). 
Careful optimisation of the SPDC spatial mode (Fig. \ref{fig:setupV61} \textbf{b}-\textbf{d}) and signal path transmission yields a heralding efficiency of $\eta_\mathrm{herald} = 0.22$. 
$\eta_\mathrm{herald}$ is the probability that a single photon is sent into the memory given a herald detection event (see Supplementary Information). 
As shown in Fig. \ref{fig:setupV61} \textbf{a} and \textbf{g}, we operate the memory in a feed-forward configuration (Methods), 
whereby herald detection triggers the preparation of the memory control field by pulse picking with a Pockels cell.
Spectrally filtering the idler projects the signal photon's marginal spectrum into Raman resonance with the control \cite{Loudon:2000}. 
Thus herald detection ensures that storage and retrieval are only attempted when a spectrally-matched photon is sent into the memory.
To benchmark the storage of heralded single photons we also store coherent states with average photon numbers similar to $\eta_{\text{herald}}$. These are generated by phase modulation of a weak control beam pick-off  \cite{Reim2010, Reim2011, England2012}. 
After the Cs cell memory, the control field is removed from the signal by polarisation and spectral filtering. 
The emerging signal is then split 50:50 on a beam splitter and directed to a pair of photon-counting detectors ($\text{D}_\text{H}$, $\text{D}_\text{V}$). See Methods and Supplementary Information for more details.

The memory efficiency and noise are determined by measuring the arrival time histograms of signal photons transmitted through and retrieved from the memory. As shown in Fig. \ref{fig:Store_retrieve_traces}, the memory performance is nearly identical for both single-photon and coherent-state inputs, with an overall efficiency $\eta_\text{tot}$ of $21$\% for heralded single photons and $29$\% for coherent states, limited by the 
control pulse energy, mode-matching between signal and control, and optical pumping (Supplementary Information). 
In addition to the memory signals, there is noise present in the read-in and the read-out time bins. 
It is visible as a pulse emitted from the memory, when only the control field is applied, which contains on average 
$\epsilon_\text{in} = \left( 6 \pm 2 \right) \cdot 10^{-2}$ photons per read-in control cycle and $\epsilon_\text{out} = \left( 15 \pm 5 \right) \cdot 10^{-2}$ photons per read-out control cycle.
The noise originates from four-wave-mixing (FWM), which is a two-step process.
First, the control generates spurious spin-wave excitations by spontaneous anti-Stokes scattering (Fig. \ref{fig:Store_retrieve_traces} \textbf{f}, Supplementary Information). 
Second, subsequent retrieval of these excitations 
by the control leads to the emission of Stokes photons into the signal mode, which cannot be separated from the signal.
The FWM mechanism causes higher noise in the read-out time bin, where it contains contributions from the read and the write control pulses. 

The signal-to-noise ratio (SNR), i.e. the ratio of the number of retrieved signal photons to the number of noise photons, is
$\mathrm{SNR}= \eta_\mathrm{tot}\eta_\mathrm{herald} / \epsilon_{\text{out}} = 0.3 \pm 0.1$,
Notably, the heralding efficiency $\eta_\text{herald}$ 
sets tight tolerances on the unconditional noise floor of the memory. 
For future implementations it is thus vital to understand the effects of noise on the properties of stored single photons. 

\begin{figure}
\begin{center}
\includegraphics[width=1\columnwidth]{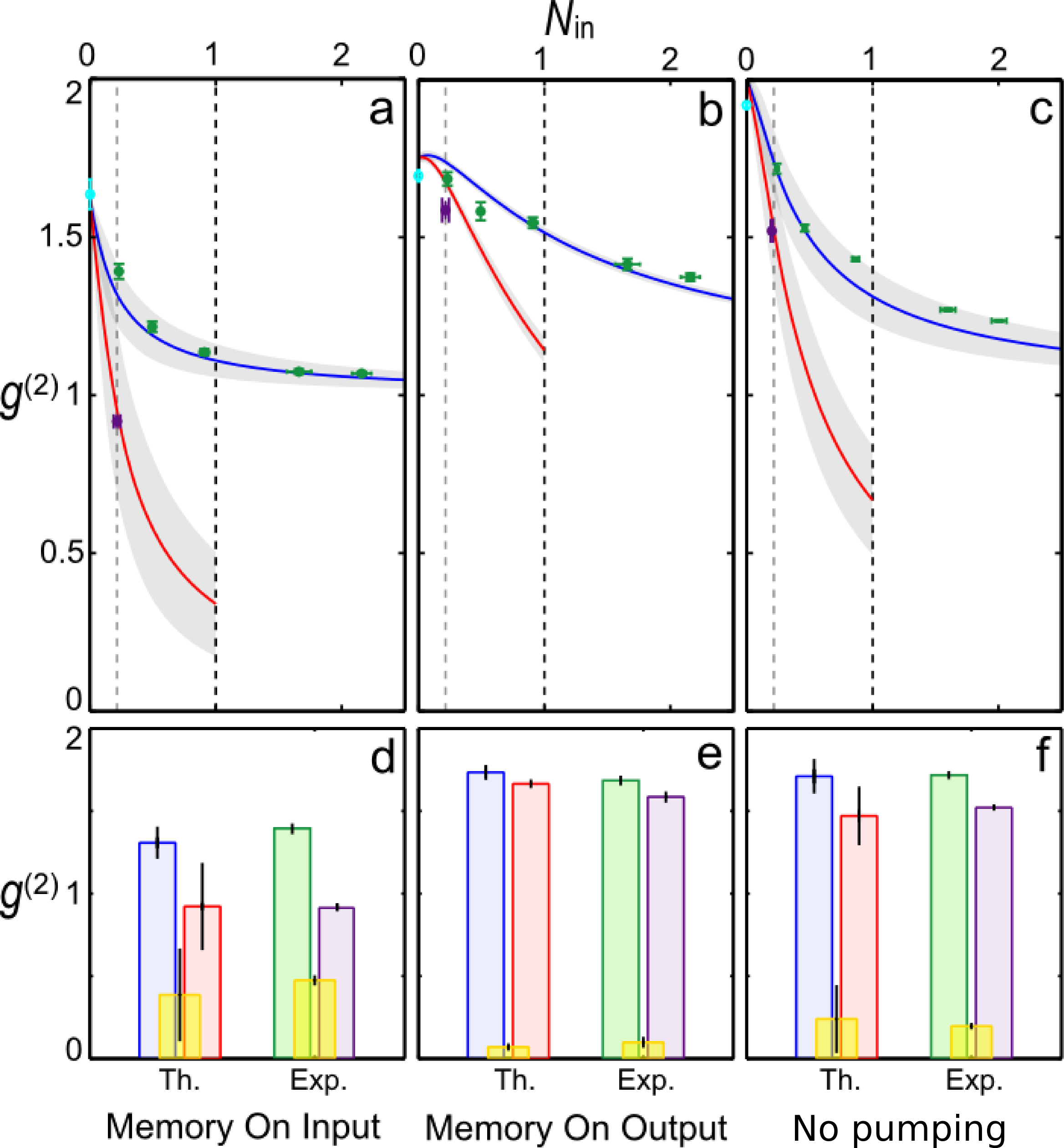}
\caption{{\bf Photon statistics measurements.} 
$g^{(2)}$ for increasing average photon number per pulse, $N_\mathrm{in}$, of the input signal in three cases:
{\bf a} not-stored signal transmitted through the memory (`Memory on input'), 
{\bf b} signal retrieved from the memory (`Memory on output'), 
{\bf c} signal transmitted through the memory without optical pumping, i.e. without storage and retrieval (`No pumping'). 
Green points are coherent-state input signal data; purple datapoints represent heralded single-photon input ($N_\text{in}=\eta_\text{herald}$), cyan points are FWM noise. Error bars derive from Poissonian counting statistics. 
Blue and Red lines show the theoretical predictions (main text, Supplementary Information) for coherent-state and heralded single-photon inputs, respectively. Shaded regions denote their standard deviations under Monte-Carlo variation of the model parameters (Supplementary Information). 
Vertical grey and black dashed lines indicate $N_\text{in}=\eta_\text{herald}$ and $N_\text{in}=1$ (perfect heralding efficiency), respectively. 
The bar plots {\bf d}-{\bf f} compare model prediction (Th.) with measurement (Exp.) for $N_\mathrm{in} \approx \eta_{\text{herald}}$,  
with equal colour coding. Yellow bars denote the $g^{(2)}$-differences between coherent-state 
and heralded single-photon inputs.
In all three cases, the experimentally observed drop in $g^{(2)}$ between coherent-state and heralded-single photon inputs is well predicted by the model.}
\label{fig:g2_plots_vertical}
\end{center}
\end{figure} 

We characterise these effects by measuring the $g^{(2)}$ autocorrelation of all field combinations shown in Fig. \ref{fig:Store_retrieve_traces} for both time bins.
From the detector count statistics, the normalised autocorrelation, conditional on herald detection, is given by $g^{(2)} = p_\mathrm{her,H,V}/ \left( p_\mathrm{her,H} \cdot p_\mathrm{her,V} \right)$  \cite{Loudon:2000} (Supplementary Information). Here, $p_{\mathrm{her, H/V}}$ is the probability that the heralding detector  ($\text{D}_\text{T}$) and one of the signal detectors ($\text{D}_\text{H}$ or $\text{D}_\text{V}$) detect a photon in coincidence. $p_{\mathrm{her, H, V}}$ is the probability for a coincidence event between all three detectors ($\text{D}_\text{T}$, $\text{D}_\text{H}$ and $\text{D}_\text{V}$).

Blocking the control field, we measure $g^{(2)} = 1.01 \pm 0.01$ for coherent-state inputs (average over all input photon numbers $N_\text{in}^{\text{COH}}$) and $g^{(2)} =  0.016 \pm 0.004$ for the single-photon input, which are close to the ideal values of $1$ and $0$. When the input signal is blocked, we measure $g^{(2)}=1.62 \pm 0.04$ and $g^{(2)}=1.70 \pm 0.02$ for the noise in the input and output time bins, respectively. With normal memory operation, i.e. signal and control applied, the photon statistics are modified by the accompanying FWM process. The noise generated by this mechanism increases the $g^{(2)}$ of the transmitted and retrieved signals, yielding the results displayed in Fig. \ref{fig:g2_plots_vertical}.
In the input time bin, transmitted coherent-state signals converge towards the ideal 
$g^{(2)}=1$ for large input photon numbers of $N_\mathrm{in} \sim 2.5$, as the SNR improves. Heralded single photons show $g^{(2)}=0.92 \pm 0.02$, slightly below the classicality boundary. Looking at the memory read-out (Fig. \ref{fig:g2_plots_vertical} \textbf{b}), we find that 
coherent states with an input photon number of $N_\text{in}^\text{COH}=  0.23 \approx \eta_{\text{herald}}$ have $g^{(2)}=1.69 \pm 0.02$ and are thus indistinguishable from the noise. Heralded single photons however show $g^{(2)}=1.59 \pm 0.03$, which is a drop in $g^{(2)}$ by more than $3$ standard deviations compared to coherent states and noise (Supplementary Information). Thus, although the $g^{(2)}$ is above the classical boundary, 
the lower value of $g^{(2)}$ for single photon input compared to that for weak coherent states
reveals the influence of the non-classical SPDC input photon statistics in the memory read-out.

The measured photon statistics are in agreement with a theoretical model of FWM that takes into account the full coherent off-resonant interaction between the incident light fields and the spin-wave excitation in the Cs ground states, with no free parameters (Fig. \ref{fig:g2_plots_vertical} \textbf{a}-\textbf{c}) (Supplementary Information). To test the model prediction further, we also investigate the statistics without spin-wave storage by preventing atomic state preparation (diode laser blocked, Fig. \ref{fig:g2_plots_vertical} \textbf{c}). In all cases the model and data are in agreement, illustrated by the bar graphs in Fig. \ref{fig:g2_plots_vertical} \textbf{d}-\textbf{f}. Notably, the same cannot be obtained by incoherently combining the noise statistics with those of the input signal fields (\cite{goldschmidt2013mode}, Supplementary Information). This close description supports the conclusion that FWM is the only remaining challenge for full photon statistics conservation during memory storage.

%
%
%

%
In this letter we have investigated the storage and retrieval of broadband heralded single photons in a room-temperature memory. 
Our results show the suitability of our system for temporal-multiplexing of heralded single photons, which is a key goal for optical quantum information science.
The detailed understanding of the FWM noise influence on the stored and retrieved signal is a critical insight for the mitigation of this remaining  challenge \cite{Lauk:2013}.
In principle, FWM can be suppressed by terminating the control field coupling to the initial atomic ground state via polarisation selection rules. This approach is ineffective in alkali vapours far from resonance \cite{Vurgaftman:2013aa}. 
As FWM is constrained by phase-matching similar to SPDC \cite{Kumar1994}, an alternative could be to introduce dispersion between the Stokes and anti-Stokes frequencies, or to use a storage medium with larger Stokes shifts \cite{England:2013aa}.
For our system, a promising approach is to reduce the density of states at the anti-Stokes frequency by placing the memory inside a low-finesse cavity or photonic-bandgap structure \cite{Sprague:2014aa}. A FWM suppression by a factor of $\sim 2.5$, achievable with these techniques, would preserve the nonclassical signature of retrieved heralded single photons (Supplementary Information). 
The resulting control over FWM will enable operation of our system as a noise-free memory for scalable photonics in the near future.

{\large\textbf{Methods}}\\
{\small
\textbf{The Raman memory protocol:}
As shown in Fig. \ref{fig:setupV61} \textbf{a} and \textbf{e}, storage of a signal field, which is blue-detuned by $\Delta = 15.2$~GHz from the Cs $\text{D}_{2}$-line, is triggered by a co-propagating, orthogonally polarised control pulse tuned into two-photon resonance with the ground-state hyperfine transition ($\Delta_\mathrm{s}=9.2\,\text{GHz}$). Signal storage transfers Cs atoms, initially prepared in the $6^{2} \text{S}_{\frac{1}{2}} F=4$ hyperfine ground state ($\ket{1}$) by optical pumping with a counter-propagating resonant diode laser, to the $6^{2}\text{S}_{\frac{1}{2}} F=3$ hyperfine ground state ($\ket{3}$), exciting a spin-wave coherence between both states \cite{Fleischhauer2000,Reim2010,Nunn2007}. Re-application of the control field drives the reverse process, producing a read-out signal and returning the atoms to $\ket{1}$. 

\textbf{Memory and signal preparation:}
As illustrated in Fig. \ref{fig:setupV61} \textbf{a}, a mode-locked Ti:Sa oscillator (\textit{Spectra Physics Tsunami}), emitting an $80\, \text{MHz}$ train of $360 \, \text{ps}$ pulses at $852 \, \text{nm}$ with $1.2 \, \text{W}$ average power (IR), is frequency doubled in a $2 \, \text{mm}$ PPKPT crystal. 
The produced second-harmonic (SHG) radiation, at $426 \, \text{nm}$, pumps SPDC in a $3 \, \mu \text{m} \times  6\, \mu \text{m} \times  2\, \text{cm}$ PPKTP waveguide (waveguide coupling efficiency $\eta_\text{C}\approx 10 \, \%$) with an average input power of $\approx 1 \, \text{mW}$. The remaining, un-converted Ti:Sa pulses, at $852\,\text{nm}$, are recovered to provide the memory control pulses by separation from the SHG on a dichroic mirror. The control field is generated by picking two consecutive pulses from the Ti:Sa pulse train for $12.5 \, \text{ns}$ storage time, which is done by a Pockels cell (PC), triggered on SPDC herald detection events (Supplementary Information). 
Coherent-state input signal pulses are produced by re-directing a small fraction of the picked control beam through an electro-optic modulator (EOM), generating $9.2 \, \text{GHz}$ sidebands on the first (write) pulse. An air-spaced etalon ($38.86\,\text{GHz}$ free-spectral range) is used to isolate the red-detuned sideband required for two photon resonance. \newline
\noindent The memory is prepared in the $6^2 \text{S}_{\frac{1}{2}} F=4$ hyperfine manifold by optical pumping with a CW frequency-stabilised external cavity diode laser. 
The beam, with $\approx 3 \, \text{mW}$ average power, is counter-propagating along the control beam path and resonant with the $6^2 \text{S}_{\frac{1}{2}} F=3$ $\leftrightarrow$  $6^2$P$_{\frac{3}{2}}$ $\text{D}_2$-transition (the upper hyperfine structure is not resolved due to Doppler broadening).

\textbf{Single-photon source and feed-forward operation:}
The PPKTP waveguide (\textit{AdvR}) produces near-degenerate, orthogonally polarised signal and idler photon pairs at $852 \, \text{nm}$ with a phasematching bandwidth of $\sim0.1 \, \text{nm}$ (Fig.~\ref{fig:setupV61}).
The idler is frequency-filtered using a $0.01 \, \text{nm}$ bandwidth volume holographic grating (\textit{ONDAX}), followed by four air-gap Fabry-Perot (FP) etalons with a free-spectral range (FSR) of $18.2\, \text {GHz}$ (two) and $103\, \text {GHz}$ (two), respectively (herald filter in Fig. \ref{fig:setupV61} \textbf{a}). 
Setting the filter resonance to $\Delta = +24.4 \, \text{GHz}$ detuning from the $6^2 \text{S}_{\frac{1}{2}} F=3 \rightarrow 6^2 \text{P}_{\frac{3}{2}}$ transition projects the SPDC signal into two-photon resonance with the control (Fig.~\ref{fig:setupV61} \textbf{e}).
Electronic signals from herald detection on a single-photon counting module (\textit{Perkin Elmer}, SPCM $\text{D}_{\text{T}}$ in Fig. \ref{fig:setupV61} \textbf{a}) are fed to the PC for control pulse picking (Fig. \ref{fig:setupV61} \textbf{g}). The heralded SPDC signal photons meanwhile propagate in an $83\,\text{m}$ long single-mode fibre (SMF), which provides sufficient time delay to compensate the electronic delays of the SPCM and the PC. A digital delay generator (\textit{SRS DG535}) provides fine adjustment of the PC trigger to synchronise the arrival times of SPDC signal photons and the control read-in pulse at the memory cell (Supplementary Information). The herald detection rate is fixed between $5.3-7.3 \,\text{kHz}$ at $1\,\text{mW}$ UV pump, which is limited by the PC, leading to a detected signal-idler coincidence rate of $\approx 30 \, \text{Hz}$ (Supplementary Information).

\textbf{Signal detection}
At the memory output, the control is separated from the signal first by polarisation filtering on a polarising beam displacer (shown as a PBS in Fig. \ref{fig:setupV61} \textbf{a}, also used for insertion of the optical pump), providing control attenuation of $\ge 40 \,$dB. 
Subsequently the signal is spatially filtered using an SMF and frequency-filtered by five air-gap Fabry-Perot etalons, with $18.2 \, \text{GHz}$ FSR (three) and $103 \, \text{GHz}$ FSR (two) (signal filter in Fig. \ref{fig:setupV61} \textbf{a}).
The etalon chain suppresses the control by $90 \, \text{dB}$ with respect to the signal, which has an on-resonance transmission of $T \approx 10 \, \%$. 
Notably, the filtering also removes any collisional fluorescence noise generated by the control fields. For measurement of the heralded autocorrelation function $g^{(2)}$, the signal is split into two spatial modes on a PBS. 
Each mode is coupled into multi-mode fibres (MMF), leading to two single-photon counting modules (\textit{Perkin Elmer}, SPCMs $\text{D}_{\text{H}}$ and $\text{D}_{\text{V}}$ in Fig. \ref{fig:setupV61} \textbf{a}). 
Detection events of both SPCMs are counted by a field-programmable gate array (FPGA) in coincidence with the herald SPCM. 
Furthermore, a time-to-amplitude-converter (TAC) - multi-channel-analyser (MCA) system is used to record arrival time histograms between herald and signal photons in the horizontally-polarised detection arm. 
The histograms shown in Fig. \ref{fig:Store_retrieve_traces} are obtained by binning the time differences between detection events registered by detectors $\text{D}_{\text{T}}$ and $\text{D}_{\text{H}}$.

\section*{Acknowledgements}
We thank A. Lvovsky and C. Kupchak for helpful discussions on FWM in Raman systems. We acknowledge Nathan Langford for early contributions to the design of the source, Michal Karpinsky for assistance with the source optimisation, and Eilon Poem for comments on the manuscript. We thank Justin Spring for assistance with the FPGA. This work was supported by the UK Engineering and Physical Sciences Research Council (EPSRC; EP/J000051/1), the Quantum Interfaces, Sensors, and Communication based on Entanglement Integrating Project (EU IP Q-ESSENCE; 248095), the Air Force Office of Scientific Research: European Office of Aerospace Research \& Development (AFOSR EOARD; FA8655-09-1-3020), EU IP SIQS (600645), the Royal Society (to J.N.), EU ITN FASTQUAST (to P.S.M.), and EU Marie-Curie Fellowships (PIIF-GA- 2011-300820 to X.-M.J.; PIEF-GA-2012-331859 to W.S.K.; PIIF-GA-2013-629229 to D.J.S.). I.A.W. acknowledges an ERC Advanced Grant and an EPSRC Programme Grant.

\section*{Author contributions}
P.S.M. built the experiment and collected all data, with assistance from T.F.M.C. and D.G.E. 
M.R.S. built the oscillator lock. 
M.B., W.S.K. and K.T.K. assisted in the source construction.
W.S.K. implemented the data acquisition electronics. 
The theoretical model was developed by J.N. 
All data analysis was carried out by P.S.M., with assistance from D.J.S. and J.N. 
The experiment was designed and conceived by J.N. and I.A.W.
P.S.M., D.J.S. and J.N. wrote the manuscript, with input from all authors. 
X.M.J. prepared the manuscript figures.

\section*{Additional information}
Supplementary information accompanies this paper at www.nature.com/naturephotonics. Reprints and permission information is available online at http://npg.nature.com/reprintsandpermissions/. Correspondence and requests for materials should be addressed to P.S.M.

\section*{Competing financial interests}
The authors declare no competing financial interests.



\end{document}


\title{Supplementary Information for ``Interfacing GHz-bandwidth heralded single photons with a room-temperature Raman
quantum memory"}

\author{P.~S.~Michelberger}
\affiliation{Clarendon Laboratory, University of Oxford, Parks Road, Oxford OX1 3PU, UK}

\author{T.~F.~M.~Champion}
\affiliation{Clarendon Laboratory, University of Oxford, Parks Road, Oxford OX1 3PU, UK}

\author{M.~R.~Sprague}
\affiliation{Clarendon Laboratory, University of Oxford, Parks Road, Oxford OX1 3PU, UK}

\author{K.~T.~Kaczmarek}
\affiliation{Clarendon Laboratory, University of Oxford, Parks Road, Oxford OX1 3PU, UK}

\author{M.~Barbieri}
\affiliation{Clarendon Laboratory, University of Oxford, Parks Road, Oxford OX1 3PU, UK}

\author{X.-M.~Jin}
\affiliation{Clarendon Laboratory, University of Oxford, Parks Road, Oxford OX1 3PU, UK}
\affiliation{Department of Physics and Astronomy, Shanghai Jiao Tong University, Shanghai 200240, PR China}

\author{D.~G.~England}
\affiliation{Clarendon Laboratory, University of Oxford, Parks Road, Oxford OX1 3PU, UK}
\affiliation{National Research Council of Canada, Ottawa, Ontario K1A 0R6, Canada}

\author{W.~S.~Kolthammer}
\affiliation{Clarendon Laboratory, University of Oxford, Parks Road, Oxford OX1 3PU, UK}

\author{D.~J.~Saunders}
\affiliation{Clarendon Laboratory, University of Oxford, Parks Road, Oxford OX1 3PU, UK}

\author{J.~Nunn}
\affiliation{Clarendon Laboratory, University of Oxford, Parks Road, Oxford OX1 3PU, UK}

\author{I.~A.~Walmsley}
\affiliation{Clarendon Laboratory, University of Oxford, Parks Road, Oxford OX1 3PU, UK}

\maketitle


\section{Experimental set-up \label{supp_setup}}
\begin{figure}[h!]
\begin{center}
\includegraphics[width=0.9 \textwidth]{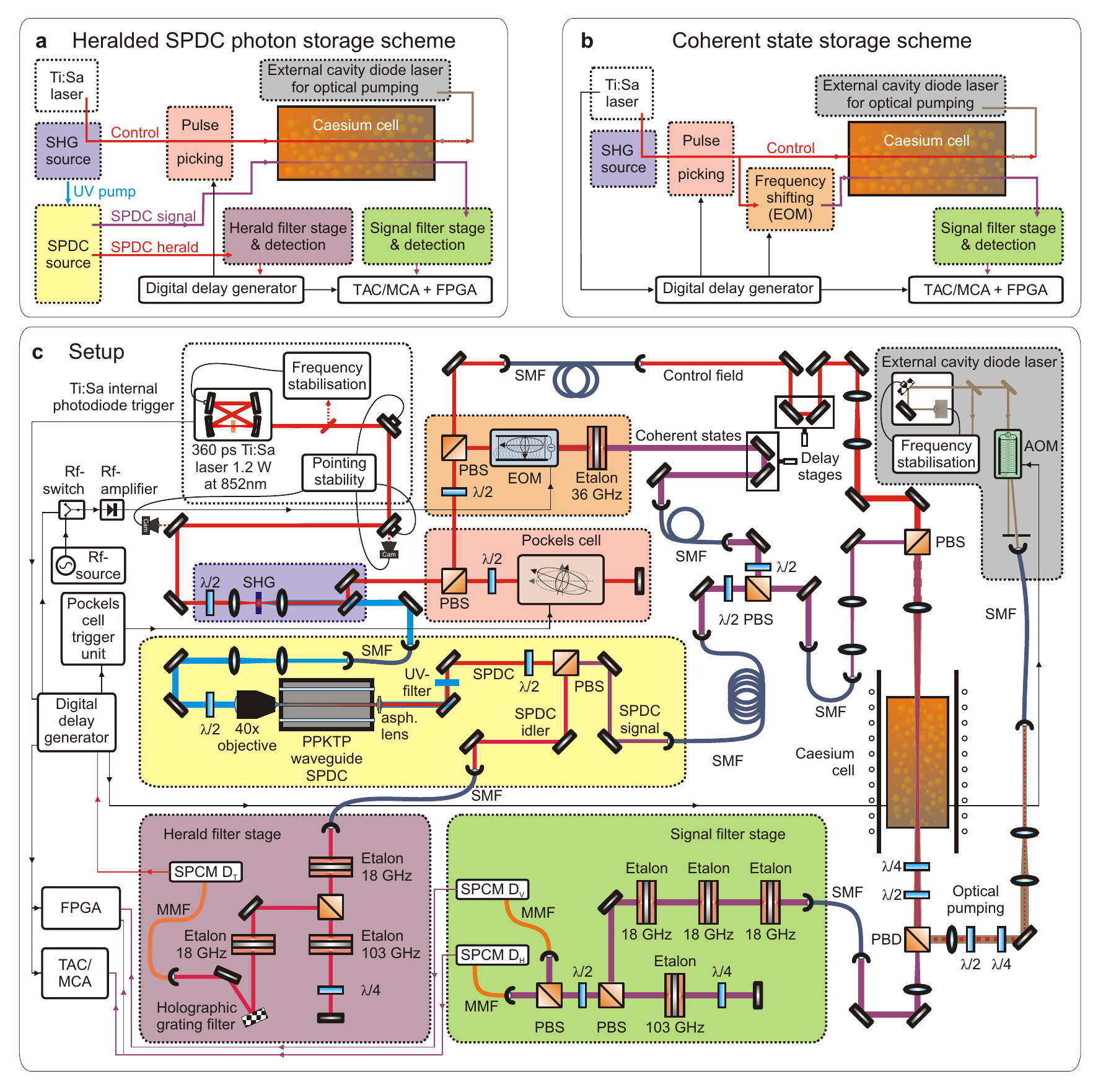}
\caption{\label{fig_supp_1} \textbf{a} Simplified configuration layout of the experimental building blocks as well as the optical and electronic signal paths for heralded SPDC photon storage by electronic feed-forward. \textbf{b} The equivalent layout for coherent state storage. \textbf{c} Schematic of the full experimental apparatus. For ease of identification, the building blocks in \textbf{a} and \textbf{b} are shaded in the appropriate colours. See text for details.}

\end{center}
\end{figure}

\noindent
Here we describe in detail the two setup configurations employed to operate the experiment with heralded single photon and coherent state inputs. 
While the system must be operated in feed-forward mode for the storage and retrieval of heralded single photons, experiments with coherent state input signals can be implemented by triggering the experiment off 
an internal photodiode of the Ti:Sa master laser \cite{Reim2011, England2012}.
The latter is significantly simpler to implement, since it does not require simultaneous operation of the SPDC source, for which reason we use this configuration for the coherent state storage measurements.
Fig. \ref{fig_supp_1} \textbf{a} and \textbf{b} contrast the differences in the components and signal paths for both triggering methods. 
Additionally, Fig. \ref{fig_supp_1} \textbf{c} illustrates the full experimental set-up, comprising all critical components for implementation of both triggering methods.

\noindent
In the case of heralded SPDC photon storage with feed-forward memory operation (Fig. \ref{fig_supp_1} \textbf{a}), the output (IR) of our frequency and beam pointing stabilised Ti:Sa laser is frequency-doubled (SHG, violet panel) with $\eta_\text{SHG} \approx 0.005 \, \frac{\%}{\text{W}}$ efficiency to generate the UV pump (blue line) for the PPKTP waveguide SPDC source (yellow panel). 
The SPDC signal (purple) and idler (red) photons are 
single-mode fibre (SMF) coupled with $\eta_\text{signal} = 72 \pm 2 \, \%$ and $\eta_\text{herald} = 73 \pm 2 \, \%$ efficiency, respectively.
Subsequently, the idler is frequency-filtered (purple panel, see Methods), with one of the etalons ($\text{FSR}=103 \, \text{GHz}$) conveniently used in double-pass (Fig. \ref{fig_supp_1} \textbf{c}).
The peak transmission of the herald filter stage is $T= 31 \pm 2 \, \%$. It is measured by $360 \, \text{ps}$ Ti:Sa laser pulses, which are resonant with the etalons 
and employed for alignment.
Idler detection on SPCM $\text{D}_\text{T}$ sends an electronic trigger pulse (thin red line Fig. \ref{fig_supp_1} \textbf{c}) to a digital delay generator unit. The digital delay generator sends appropriately delayed gating pulses the FPGA and TAC/MCA modules for coincidence counting with the signal SPCMs  $\text{D}_\text{H}$ and $\text{D}_\text{V}$, as well as the Pockels cell trigger unit (see feed-forward operation in Methods).
The unconverted Ti:Sa pulses are pulse-picked by the Pockels cell (pink panel), SMF-coupled (SMF length $10 \, \text{m}$, coupling efficiency $\eta_{\text{control}}\approx 53 \, \%$) and delayed to temporally overlap with the SPDC signal pulses in the caesium (Cs) cell (see main text and Methods for a pulse timing description).
SPDC signal photons are also delayed by two SMFs of $83 \, \text{m}$ and $7.97 \, \text{m}$ length, respectively, before being combined with the control field on a PBS in front of the Cs cell. 
In order to switch between heralded SPDC photons and coherent states as memory input signals, a PBS is used to couple the coherent state pulses into the second signal SMF.
Notably, only one input type is applied at any one time. 
Thus, for heralded SPDC storage experiments, the coherent state arm is blocked and vice versa.

\noindent
In the case of coherent state storage (Fig. \ref{fig_supp_1} \textbf{b}), the digital delay generator 
is triggered by the Ti:Sa intra-cavity photodiode signal. 
Additionally the digital delay generator also switches the EOM (see Methods), used to generate the coherent state signal by phase modulation of a small pick-off from the control pulses, which are, as before, selected from the unconverted IR output of the SHG source by the Pockels cell. 
Using a fast rf-switch ($6 \,\text{ns}$ rise time), the EOM is only supplied with the required $9.2 \, \text{GHz}$ modulation frequency for the input time bin. 
As a result, coherent state input signal pulses are only present in the read-in time bin,
whereas the control pulses for storage and retrieval exist for both
time bins \cite{Reim2011,England2012}.
Notably, switching the experimental control electronics between the configurations for heralded SPDC photon and coherent state input signals only requires a change in the settings of the digital delay generator.

\noindent
In both configurations, the signal is stored in the Cs cell, which is surrounded by a single layer magnetic shield \cite{Reim2011,England2012} (orange panel). The Cs vapour is heated to $70 \, ^{\circ} \text{C}$, with a cell cold spot at $67.5  \, ^{\circ} \text{C}$. 
The counter-propagating optical pumping beam from the frequency stabilised external cavity diode laser (grey panel) is on continuously during the heralded single photon storage experiment. 
It is turned off by an acusto-optic modulator (AOM), also gated by the digital delay generator, during the storage time for coherent state input signals. 
This difference results from the long AOM switching times in our setup configuration of $\approx 1 \, \mu \text{s}$,
which exceeds the available SMF time delay of the heralded SPDC signal photons. 
For coherent state storage this limitation does not arise.
Here, in contrast to the probabilistic occurrences of SPDC herald detection events, the Ti:Sa photodiode provides a deterministic repetition rate of $80 \, \text{MHz}$.
Thus the AOM switch-off can be postponed by one repetition cycle to turn the optical pumping off for the following storage event. To this end the $5.72 \, \text{kHz}$ repetition rate for memory experiments leaves $1.7 \, \text{ms}$ between storage events, which suffices for AOM switching.
Notably, while the presence of the optical pumping slightly reduces the overall memory efficiency $\eta_\text{tot}$ for heralded SPDC photons (see \ref{supp_memeff}), we have certified that this difference in the optical pumping timing has no influence on the photon statistics. It neither affects the $g^{(2)}$ of the retrieved signal, nor the $g^{(2)}$ of the noise in both time bins. 

\noindent
Behind the memory, signal and control are separated spatially on a calcite polarising-beam displacer (PBD), which provides high-quality polarisation extinction ($\ge 40 \, \text{dB}$) of the undesired mode. 
The signal is SMF-coupled ($\eta_\text{sig.filt.} \approx 88 \, \%$) and frequency-filtered (green panel, see Methods), again using the $\text{FSR} = 103 \, \text{GHz}$ etalon in double-pass.
The peak transmission of the signal filter for $360 \, \text{ps}$ Ti:Sa laser pulses is $T\approx 10 \, \%$, which includes SMF-coupling and the transmission of the optics behind the Cs cell. 
To measure $g^{(2)}$, the signal is split up $50:50$ on a PBS, multi-mode fibre (MMF) coupled and detected using SPCMs $\text{D}_\text{H}$ and $\text{D}_\text{V}$. 
Their electronic output pulses feed into the TAC/MCA ($\text{D}_\text{H}$) and the FPGA ($\text{D}_\text{H}$, $\text{D}_\text{V}$) for coincidence detection with events registered by the herald SPCM $\text{D}_\text{T}$.



\section{Heralding efficiency \label{supp_hereff}}
\noindent
The heralding efficiency of $\eta_{\text{herald} }= 0.22$, quoted in the main text, represents the probability of observing a signal photon at the memory input facet, upon detection of a herald photon. 
In other words, this corresponds to the number of photons sent into the memory when detecting a herald photon, which triggers a memory experiment by picking a control pulse sequence with the Pockels cell.
The heralding efficiency is recorded by turning off the memory interaction (blocking the control), while still optically pumping the Cs atoms. 
The accumulated coincidences between the herald and each of the signal detectors 
$c_\text{her,sig} = c_\text{her,H} + c_\text{her,V}$ are normalised by the number $c_\text{her}$ of counts registered on the herald detector $\text{D}_\text{T}$. This ratio is corrected for the combined transmission of the Cs cell, the optics behind the memory output and the transmission of the signal filter stage, $T_\text{tot} \approx 10 \, \%$, as well as the detection efficiency of the SPCMs $\text{D}_{\text{H}}$, $\text{D}_{\text{V}}$, which are assumed as $\eta_\text{SPCM} \approx 50 \, \%$.
From these numbers the heralding efficiency is obtained by 

$$
\eta_{\text{herald}} = \frac{c_\text{her,sig} }{c_\text{her} \cdot T_\text{tot} \cdot  \eta_\text{SPCM} }. 
$$

\noindent
Notably this definition assumes negligible contributions from higher order SPDC emissions, which is justified in our case and can be seen by the low $g^{(2)}=0.016 \pm 0.004$ measured for heralded SPDC input signals (main text).

\section{Memory efficiency \label{supp_memeff}}

\noindent
The memory efficiency is calculated from the coincidence count rates between signal and herald SPCMs for the memory input and output time bins, $c_{\text{her,H/V}}^{\text{in/out}}$. These are measured for 4 different input field combinations ($l$), referred to as experimental settings, to access all contributions to the detected signal (see also Fig.~2): 
\begin{enumerate}
\item Memory on (\textit{scd}): signal (\textit{s}), control (\textit{c}) and optical pumping (\textit{d}) are sent into the Cs cell (blue lines in Fig.~2).
\item Input signal (\textit{sd}): control is blocked (green lines in Fig.~2). 
\item Noise (\textit{cd}): input signal is blocked (red lines in Fig.~2).
\item Optical pumping background (\textit{d}), i.e. signal and control are blocked.
\end{enumerate}
For the arrival time histograms, shown in Fig.~2, the $c_{l}^{\text{in/out}}$ are the integrated pulses, 
normalised by the sum of all herald trigger events $c_\text{her}$. 
The memory efficiency is given by:

\begin{equation}
\eta_{\text{tot}}=\left[c_\textit{scd}^{\text{out}}-c_\textit{cd}^{\text{out}}-\left(c_\textit{sd}^{\text{out}}-c_\textit{d}^{\text{out}} \right) \right]/\left(c_\textit{sd}^{\text{in}} - c_\textit{d}^{\text{in}} \right),
\label{eq_mem_eff}
\end{equation}

\noindent
whereby the error on $\eta_\text{tot}$ derives from Gaussian error propagation using Poissonian count rate errors on the $c_{j}^{\text{in/out}}$. 
Notably, $c_\textit{d}^{\text{in/out}}$ is negligible and has thus been omitted in Fig.~2. 
The memory efficiency for heralded single photon input signals ($N_\text{in}=\eta_\text{herald} = 0.22$) is $\eta_{\text{tot}} = 21.1 \% \pm 1.9\%$ (see main text).  
For coherent states at similar input photon number ($N_{\text{in}}^{\text{COH}} = 0.23$ photons per pulse)  the memory efficiency is $\eta_{\text{tot}} = 29.02 \% \pm 0.9 \%$. 
The difference is firstly due to the absence of optical pumping during storage for coherent state inputs (see \ref{supp_setup}). Secondly a residual mode mismatch between the spectra of heralded SPDC photons and the control pulses also contributes, which leads to a decrease in memory read-in efficiency of SPDC signal photons compared to storage of coherent state inputs. The mismatch arises from filtering of the SPDC idler photons, which does not perfectly project the signal into the spectral mode of the control.

\section{$g^{(2)}$ measurement \label{supp_g2meas}}

\subsection{Measurement procedure for $g^{(2)}$\label{supp_g2procedure}}

\noindent
To determine the $g^{(2)}$- values, presented in Fig.~3, with sufficient precision, we need to collect count rate statistics on the coincidence probabilities $p^{\text{in/out}}_{\text{her,H/V}}=\frac{c^{\text{in/out}}_{\text{her,H/V}}}{c_{\text{her}}}$ 
and triple coincidence probabilities $p^{\text{in/out}}_{\text{her,H,V}}=\frac{c^{\text{in/out}}_{\text{her,H,V}}}{c_{\text{her}}}$ over long time scales.
Long measurement times are necessary due to a technical limitation on the experiment repetition rate $f_\text{rep}$, which arises from the Pockels cell. 
At too high a repetition rate, the Pockels cell pulse-picking extinction ratio, i.e. the amplitude ratio of unpicked pulses with respect to picked pulses, degrades, for which reason experiments are conducted with $f_\text{rep} = 5.3 - 7.3 \, \text{kHz}$.
Notably $f_\text{rep}$ equals the number of herald detection events $c_\text{her}$ and thus sets an upper boundary on the brightness requirements of the SPDC source.\\
The resulting limitation on detected coincidence counts $c^{\text{in/out}}_{\text{her,H/V}}$ and $c^{\text{in/out}}_{\text{her,H,V}}$ between the herald SPCM and the H-, V-signal SPCMs ($\text{D}_\text{T}$, $\text{D}_\text{H}$ and  $\text{D}_\text{V}$ in Fig.~1) make it necessary to run the experiment over several days for each input signal type (heralded SPDC photons, coherent states at different mean photon numbers).
Table \ref{supp:table1} shows the total measurement time $\Delta t_{\text{meas}}$ for each input signal type and the average detected coincidence count rates.
%
\begin{table}[h!]
   \centering
   \begin{tabular}{@{}| l |c |c |c |c |c |c |c |@{}} 
	\hline
      Signal type	& $\Delta t_{\text{meas}}$ [min.]  & $c^{\text{in}}_{\text{her,H}}$ [Hz] 	& $c^{\text{in}}_{\text{her,V}}$ [Hz] 	& $c^{\text{out}}_{\text{her,H}}$ 	& $c^{\text{out}}_{\text{her,V}}$ 	& $c^{\text{in}}_{\text{her,H,V}}$ 	& $c^{\text{out}}_{\text{her,H,V}}$ \\
      \hline
     	SPDC \textit{scd} $N_{\text{in}} = 0.22$	& 295	& $28.37 \pm 0.08$  & $28.6 \pm 0.08$ 	& $29.1 \pm 0.1$  	& $29.4 \pm 0.1$ 	& $0.126 \pm 0.003$ 	& $0.229 \pm 0.004$	\\
     	Coh. \textit{scd} $N_{\text{in}} = 0.23$	& 345	& $26.49 \pm 0.05$  & $25.8 \pm 0.05$ 	& $32.49 \pm 0.07$	& $32.31\pm 0.07 $	& $0.166 \pm 0.003$	& $0.309 \pm 0.004$	\\
     	Coh. \textit{scd} $N_{\text{in}} = 0.49$	& 160	& $51.91 \pm 0.09$  & $51.4 \pm 0.1$	& $34.4 \pm 0.1$	& $34.1 \pm 0.1$	& $0.567 \pm 0.008$	& $0.325 \pm 0.006$	\\
     	Coh. \textit{scd} $N_{\text{in}} = 0.91$	& 171	& $80.8   \pm 0.15$	& $80.6 \pm 0.17$	& $55.9 \pm 0.2$	& $55.5 \pm 0.2$	& $1.29 \pm 0.01$		& $0.84 \pm 0.01$		\\
     	Coh. \textit{scd} $N_{\text{in}} = 1.66$	& 116	& $127.7 \pm 0.35$	& $130 \pm 0.4$ 	& $61.7 \pm 0.2$	& $62.6 \pm 0.2$	& $3.11 \pm 0.03$		& $0.95 \pm 0.01$		\\
    	Coh. \textit{scd} $N_{\text{in}} = 2.16$	& 147	& $169.4 \pm 0.4$	& $170 \pm 0.4$	& $80 \pm 0.3$		& $80.4 \pm 0.2$	& $5.38 \pm 0.03$		& $1.54 \pm 0.02$		\\
     	Noise \textit{cd}				& 1702	& $7.58 \pm 0.02$	& $7.26 \pm 0.02$	& $19.13 \pm 0.05$	& $18.95 \pm 0.05$	& $0.017 \pm 0.0004$	& $0.12 \pm 0.001$ 		\\     						
      \hline
   \end{tabular}
   \caption{Total measurement times and observed average coincidence rates between $\text{D}_\text{T}$, $\text{D}_\text{H}$ and $\text{D}_\text{V}$, respectively . Numbers are shown for heralded SPDC (SPDC \textit{scd}) and coherent state (Coh. \textit{scd}) input signals with the memory interaction active, as well as for FWM noise only, i.e. detected noise when no signal is sent into the memory (Noise \textit{cd}).}
   \label{supp:table1}
\end{table}
\noindent
Comparability between different measurement days is ensured by following a procedure, where a set sequence of experiment runs is conducted during each day. 
In each run the $g^{(2)}$ of the signal input is measured first for $\approx 10 \, \text{min.}$ by blocking the control field (\textit{sd}).
Subsequently, the $g^{(2)}$ for the transmitted and retrieved signal with active memory interaction (\textit{scd}) 
is recorded for $\approx 30 \, \text{min}$. 
Finally, the $g^{(2)}$ of the FWM noise is measured for $\approx 30 \, \text{min.}$ by blocking the signal input to the memory (\textit{cd}).\\
For each run, we also calculate the memory efficiency using Eq. \ref{eq_mem_eff}.
Notably the counts $c_{l}^{\text{in/out}}$, entering Eq. \ref{eq_mem_eff}, are scaled according to the measurements times of each setting $l$. 
The quoted efficiencies $\eta_{\text{tot}}$ (main text and \ref{supp_memeff}) represent the average memory efficiencies over all runs.\\
Between each measurement run, the spatial overlap between signal, control and optical pumping is optimised for maximum memory efficiency. Furthermore, the filter stage alignment as well as the source heralding efficiency are inspected and re-optimised, if required. 
This procedure minimises systematic drifts in the experimental apparatus.


\subsection{Calculation of $g^{(2)}$\label{supp_g2calc}}
\noindent
In the calculation of the $g^{(2)}$-values for all input signal types, as shown by the datapoints in Fig.~3, with the exception of the input $g^{(2)}$ for coherent states,
we use Poissonian counting statistics and sum the observed count rates over all measurement runs $j$. 
Consequently the herald counts $c_\text{her}$, the coincidence counts $c^{\text{in/out}}_{\text{her},H/V}$ 
as well as the triple coincidence counts $c^{\text{in/out}}_\text{her,H,V}$ 
are given by 
$c^{t}_{k,l} = \sum_{j=1}^{N_{r}} c^{t}_{j,k,l}$, 
with $k \in \{ \text{(her)},\text{(her,H)},\text{(her,V)},\text{(her,H,V)} \}$ and the total number of measurement runs $N_r$ for each setting $l \in \{ \text{\textit{scd}}, \text{\textit{sd}}, \text{\textit{cd}} \}$ and time bin $t \in \{ \text{in}, \text{out} \}$. 
The detection probabilities $p^{t}_{k,l} = \frac{c^{t}_{k,l}}{c_{\text{her},l} }$, with $k \in \{ \text{(her,H)},\text{(her,V)},\text{(her,H,V)} \}$, and in turn also 
\begin{equation}
g^{(2)}_{l,t}= \frac{p^{t}_{\text{(her,H,V)},l}}{p^{t}_{\text{(her,H)},l}\cdot p^{t}_{\text{(her,V)},l}},
\label{eq:supp1}
\end{equation}
\noindent
are obtained using the summed coincidence counts $c^{t}_{k,l}$. 
The errors are given by Gaussian error propagation, using the Poissonian errors for the individual coincidence counts $\Delta c^{t}_{k,l} = \sqrt{c^{t}_{k,l}}$. 
This procedure is justified as the observed count rates for the each type of input signal (i.e. heralded SPDC or coherent state with fixed photon number) are similar for each measurement day. 

\noindent
We test the validity of this procedure by also determining the $g^{(2)}_{j,l,t}$ values obtained for each measurement run $j$,
by 
\begin{equation}
g^{(2)}_{j,l,t}= \frac{c^{t}_{j,\text{(her,H,V)}} \cdot c_{j,\text{her}}}{c^{t}_{j,\text{(her,H)}} \cdot c^{t}_{j,\text{(her,V)}}}.
\label{eq:supp2}
\end{equation}
Performing a double-sided, one-sample Student T-test on the $g^{(2)}_{j,l,t}$ with $g^{(2)}_{l,t}$ from Eq. \ref{eq:supp1} as the assumed population mean, yields that the Null hypothesis ($H_0$) of $g^{(2)}_{l,t}$ being the mean of the $g^{(2)}_{j,l,t}$ cannot be rejected for any input signal type $l$ and time bin $t$ with $\ge 95 \, \%$ confidence. 
The applicability of the T-test is also tested by performing a Shapiro-Wilk test on the $g^{(2)}_{j,l,t}$, where the Null hypothesis of the $g^{(2)}_{j,l,t}$ being normally distributed cannot be rejected with $ \ge 95 \, \%$ confidence for all input signal types $l$ and time bins $t$. 
Notably, a normal distribution of the $g^{(2)}_{j,l,t}$ is expected from the central limit theorem, despite the fact that the variables $c^{t}_{j,k,l}$, going into Eq. \ref{eq:supp1}, are count rates with Poissonian distributions \cite{Loudon:2000}.

\noindent
To obtain $g^{(2)}_\text{in} = 1.01 \pm 0.01$ for coherent states inputs, i.e. the measured $g^{(2)}$ when the control field is blocked (\textit{sd}), we combine the data for all input photon numbers $N_\text{in}^{\text{COH}}$ (green points in Fig.~3 and Fig. \ref{fig:supp_g2_plots_vertical}) and calculate a weighted average over the $g^{(2)}_{l,t}$, with the total measurement time $\Delta t_\text{meas}$ for each input photon number $N_\text{in}^{\text{COH}}$ as a weighting factor. 
Using the weighted mean over the $g^{(2)}_{j,l,t}$ for all individual measurement runs $j$, instead of the sum over all $j$, is required as the $c^{t}_{j,k,l}$ differ between the different input photon numbers $N_\text{in}^{\text{COH}}$ (see table \ref{supp:table1}).

\subsection{Statistical significance of $g^{(2)}$ difference between heralded SPDC photons and coherent states\label{supp_g2significance}}

\noindent
To evaluate the statistical significance of the drop in $g^{(2)}$ between coherent states at $N_\text{in}^{\text{COH}} = 0.23$ and heralded single photons at $\eta_\text{herald}$ (Fig. ~3) retrieved from the memory, we perform a one-sided, two-sample Welch test on the individual $g^{(2)}_j = g^{(2)}_{j,\textit{scd},\text{out}}$ for both input signal types 
(with the measurement runs $j$ as in \ref{supp_g2procedure}).
A Welch test is chosen since the $g^{(2)}_j$ for coherent states and SPDC signal photons have unequal sample sizes, are drawn from different populations and have different variances. 
Testing the Null hypothesis ($H_0$) that the $g^{(2)}_j$-samples for coherent states and heralded single photons have the same population mean, we obtain a rejection of the $H_0$ with a confidence level of $\ge 99.7 \, \%$ (p-value $=8.7 \cdot10^{-4}$), which corresponds to a significance of $\ge 3$ standard deviations. \\
Similar results are obtained when replacing the $g^{(2)}_j$ of the coherent state signal with those of the FWM noise.
To complete the argument we furthermore test the $g^{(2)}_j$ for coherent states at $N_\text{in}^{\text{COH}} = 0.23$ against those of the FWM noise, again under $H_0$ that the population means are equal. 
In this case, we cannot reject $H_0$ with any reasonable level of confidence (p-value $=0.967$). 

\noindent
In conclusion, we can thus state that there is a statistically significant difference between the $g^{(2)}_j$-values of retrieved heralded single photons with respect to those of retrieved coherent states at equal photon number, as well as with respect to the $g^{(2)}_j$ of the FWM noise. 
In contrast, there is no significant difference between the coherent state and FWM $g^{(2)}_j$-values.

\section{Modelling the $g^{(2)}$ autocorrelation measurements\label{supp_g2model}}

\noindent
\subsection{Coherent $g^{(2)}$ model\label{supp_g2cohmodel}}
\noindent
Fig.~3 in the main text shows our measurements of the $g^{(2)}$ field autocorrelations alongside the predictions of a theoretical model that includes the effects of four-wave mixing seeded by spontaneous anti-Stokes scattering. The agreement is extremely good, which suggests that four-wave mixing is the only significant source of noise in the memory. Here we briefly outline the theoretical model, which is based on \cite{Reim2011_supp, England:2013aa}, and is in fact identical to \cite{Wu:2010aa}. We consider one-dimensional propagation along the $z$-axis, normalised so that $z$ runs from 0 to 1, of Stokes (signal) and anti-Stokes fields $S$, $A$ through an ensemble of three-level $\Lambda$-type atoms (Fig.~1 \textbf{e}, \textbf{f}) in the presence of a control pulse with time-dependent Rabi frequency $\Omega(\tau)$, where $\tau = t-z/c$ is the local time, at time $t$, in a frame propagating with the pulse at velocity $c$. After adiabatic elimination of the excited state $\ket{2}$ \cite{Raymer:1981aa,Gorshkov:2007aa}, the Maxwell-Bloch equations describing the Raman interaction of the fields with the ground state coherence are given by
\begin{eqnarray}
\nonumber [\partial_z + \mi \kappa ]S&=&\mi C B,  \\
\nonumber \partial_z A^\dagger &=&-\mi C'B, \\
\label{Maxwell} [\partial_\epsilon + \mi  s] B &=& \mi w[CS +  C'A^\dagger],
\end{eqnarray}
where $B = \sum_{j:\mathrm{atoms\,at}\,z} \ket{3_j}\!\!\bra{1_j}e^{\mi k_B z}$ is the annihilation operator for spin wave excitations with wavevector $k_B = k_A-k_S$, with $k_{S,A}$ the Stokes, anti-Stokes wave vectors. The \emph{effective time} $\epsilon = \epsilon(\tau) = \alpha \int_{-\infty}^\tau |\Omega(\tau')|^2\,\rmd \tau'$, with $\alpha$ such that $\epsilon(\infty)=1$, parameterises the adiabatic following of the control pulse \cite{Nunn2007}. With this coordinate transformation, the dynamic Stark shift is $s = (\alpha \Delta)^{-1} + (\alpha \Delta')^{-1}$, where $\Delta' = \Delta + \Delta_\mathrm{s}$ is the anti-Stokes detuning (Fig.~1 \textbf{e}). The dimensionless coupling constant $C$ is given by $C^2 = d\gamma/ \alpha \Delta^2$, where $d$ is the resonant optical depth and $\gamma$ the homogeneous linewidth of the excited state \cite{Gorshkov:2007aa,Reim2010}. $C'$ is identical to $C$, except that $\Delta$ is replaced by $\Delta'$. The population inversion is $w = p_1-p_3$, where $p_1=\av{\outprod{1}{1}}$ and $p_3=\av{\outprod{3}{3}}$ are the initial occupation probabilities of the ground states $\ket{1}$, $\ket{3}$, respectively. Finally, we have defined the four-wave mixing phase mismatch $\kappa = 2k_\Omega - k_S-k_A$, where $k_\Omega = L\omega_\mathrm{c}/c + d\gamma (p_3/\Delta + p_1/\Delta')$ is the wavevector of the control with frequency $\omega_\mathrm{c}$. We also have $k_S = L\omega_S/c + d\gamma[p_1/\Delta + p_3/ (\Delta-\Delta_\mathrm{s})]$, and $k_A = L\omega_A/c + d\gamma[p_3/\Delta' + p_1/ (\Delta+2\Delta_\mathrm{s})]$. Here the length $L$ of the cell appears, to account for the normalisation of the $z$-coordinate. These expressions are simply derived by considering the refractive index for each field, due to off-resonant interaction with the atomic transitions. Note that on Raman resonance we have $\omega_S = \omega_\mathrm{c} - \Delta_\mathrm{s}$ and $\omega_A = \omega_\mathrm{c} + \Delta_\mathrm{s}$. Our experiments were done with a 7.5~cm caesium cell held at $70^\circ$~C, giving a resonant optical depth $d \approx 1800$. We focus 10~nJ control pulses with duration $T_\mathrm{c}=360$~ps into the cell with a waist $\sim$300~$\mu$m, giving a peak Rabi frequency $\Omega_\mathrm{max}=4.2$~GHz, so that $\alpha = 0.31$~ns. We operate blue-detuned from resonance with the $D_2$ line, with a detuning of $\Delta = 15.2$~GHz, and the Stokes shift in caesium is $\Delta_\mathrm{s} = 9.2$~GHz. Finally, the homogeneous linewidth is $\gamma = 16$~MHz, so that we have $C = 0.82$. When the memory is operated normally, the atoms are optically pumped into the ground state $\ket{1}$, and we set $p_3 = 0$ ($w = 1$). For the case where the optical pumping beam is blocked, the atoms thermalise and we set $p_3 = 0.5$ ($w=0$).

\noindent
The system of coupled partial differential operator equations (\ref{Maxwell}) is linear, and the solutions can therefore be written as a linear mapping from initial input to final output fields via the system's Green's functions $G_{jk}$. For example, the signal field transmitted through the memory during the storage interaction is
\begin{equation}
\label{Sin}
S_\mathrm{trans}(\epsilon) = \int_0^1 G_{SS}(\epsilon,\epsilon')S_\mathrm{in}(\epsilon')\,\rmd\epsilon' + \int_0^1 G_{AS}(\epsilon,\epsilon')A^\dagger_\mathrm{vac}(\epsilon')\,\rmd\epsilon' + \int_0^1 G_{BS}(\epsilon,z)B_\mathrm{therm}(z)\,\rmd z,
\end{equation}
where `vac' and `therm' denote initial vacuum and thermal states for the anti-Stokes and spin-wave fields. The signal field retrieved from the memory after a storage time $T$ is given by a similar expression,
$$
S_\mathrm{ret}(\epsilon) = \int_0^1 G_{SS}(\epsilon,\epsilon')S_\mathrm{vac}(\epsilon')\,\rmd\epsilon' + \int_0^1 G_{AS}(\epsilon,\epsilon')A^\dagger_\mathrm{vac}(\epsilon')\,\rmd\epsilon' + \int_0^1 G_{BS}(\epsilon,z)B_T(z)\,\rmd z,
$$
where to a good approximation in our experiment we may neglect decoherence and set $B_T = B_\mathrm{out}$, since $T=12.5$~ns, while the memory lifetime is $>1$~$\mu$s \cite{Reim2011}). Here, the spin-wave at the end of the storage interaction is given by
$$
B_\mathrm{out}(z) =  \int_0^1 G_{SB}(z,\epsilon)S_\mathrm{in}(\epsilon)\,\rmd\epsilon + \int_0^1 G_{AB}(z,\epsilon)A^\dagger_\mathrm{vac}(\epsilon)\,\rmd\epsilon + \int_0^1 G_{BB}(z,z')B_\mathrm{therm}(z')\,\rmd z'.
$$
The $g^{(2)}$ autocorrelation for a short-pulsed time-dependent field measured by slow detectors is given by
\begin{equation}
\label{g2_exp}
g^{(2)}_\mathrm{x} = \frac{ \int \!\!\!\int \rmd t\,\mathrm{d} t' \langle S_\mathrm{x}^\dagger(t)S_\mathrm{x}^\dagger(t')S_\mathrm{x}(t')S_\mathrm{x}(t)\rangle }{ \left[\int \mathrm{d} t'' \langle S_\mathrm{x}^\dagger(t'')S_\mathrm{x}(t'')\rangle \right]^2 } = \frac{ \int \!\!\!\int \rmd \epsilon\,\mathrm{d} \epsilon' \langle S_\mathrm{x}^\dagger(\epsilon)S_\mathrm{x}^\dagger(\epsilon')S_\mathrm{x}(\epsilon')S_\mathrm{x}(\epsilon)\rangle }{ \left[\int \mathrm{d} \epsilon'' \langle S_\mathrm{x}^\dagger(\epsilon'')S_\mathrm{x}(\epsilon'')\rangle \right]^2 },
\end{equation}
where the label `x' is either `ret' or `trans'. Since the initial field operators satisfy bosonic commutation relations (or in the case of $\av{[B_\mathrm{in}(z),B_\mathrm{in}^\dagger(z')]}=w\delta(z-z')$, \emph{boson-like}) and their expectation values on the initial state of the atomic-optical system are known, the expectation values of normally ordered products of the output field operators can be computed from products of the Green's functions \cite{Reim2011_supp,Wu:2010aa}, enabling the calculation of the $g^{(2)}$. The Green's functions can be found analytically \cite{Wu:2010aa}, although for convenience we use a previously-developed numerical code. Full details will be given elsewhere. However here we note that the $g^{(2)}$ predicted by this calculation depends on the statistics of the input field, and on its brightness, through taking expectation values involving the field operator $S_\mathrm{in}$ in Eq.~(\ref{Sin}).
\subsection{Monte-Carlo error propagation}
\label{monte}
We used a Monte-Carlo approach to generate the shaded error regions around the theoretical predictions plotted in Fig.~3 in the main text, and in Fig.~\ref{fig:supp_g2_plots_vertical} below. We computed the theory predictions 1000 times, with each input parameter drawn from a Gaussian distribution with standard deviation set by its experimental uncertainty. For each value of $N_\mathrm{in}$, the standard deviations of the predictions were then used to set the vertical width of the error regions in the plots. The errors on the various input parameters were estimated as follows.
\begin{itemize}
\item[] $\mathrm{error}(d) = 100$,
\item[] $\mathrm{error}(\gamma) = 0.5$~MHz,
\item[] $\mathrm{error}(\Delta) = 200$~MHz,
\item[] $\mathrm{error}(T_\mathrm{c}) = 20$~ps,
\item[] $\mathrm{error}(\Omega_\mathrm{max}) = 100$~MHz.
\end{itemize}

\subsection{Incoherent $g^{(2)}$ model\label{supp_g2incohmodel}}
\noindent
We cross-check our description for the combined dynamics of the signal and the FWM noise by investigating the expected photon statistics using an incoherent interaction model. 
In this case, the combined photon statistics $\left( g^{(2)}_\text{tot} \right)$ is assumed to consist of a superposition between FWM noise $\left( g^{(2)}_\text{noise} \right)$ and the input signal $\left( g^{(2)}_\text{sig} \right)$, whereby both fields are imagined to be combined on a beam-splitter. 
The expected value for $g^{(2)}_\text{tot}$ is derived following the argumentation presented in \cite{goldschmidt2013mode}, providing

\begin{equation}
g^{(2)}_\text{tot} = \frac{ \left(N_\text{sig} \right)^2 \cdot g^{(2)}_\text{sig} + 2 N_\text{sig} \cdot N_\text{noise} + \left( N_\text{noise} \right)^2 \cdot g^{(2)}_\text{noise} }{\left(N_\text{sig}+ N_\text{noise} \right)^2},
\label{eq_goldschmidt}
\end{equation}

\noindent
which depends on the number of signal photons, $N_\text{sig}$, and noise photons, $N_\text{noise}$, per pulse contributing to the mixture. 
In the memory on cases, $N_\text{sig}$ is either the transmitted (leakage) fraction of the signal, 
$N_{\text{sig}}= (1-\eta_{\text{read-in}} ) \cdot N_\text{in}^{\text{\textit{sd}}} $, 
or the retrieved fraction of the signal, 
$N_{\text{sig}}= \eta_{\text{tot}} \cdot N_\text{in}^{\text{\textit{sd}}}$. 
Here $N_\text{in}^{\text{\textit{sd}}}$ is the input photon number for the respective signal type
(see \ref{supp_memeff}), 
$\eta_\text{read-in}$ is the memory read-in efficiency and $\eta_{\text{tot}}$ is the total memory efficiency (storage and retrieval).
For the example of heralded single photons, these numbers are $N_\text{in}^{\text{\textit{sd}}} = \eta_\text{herald} = 0.22 \pm 0.03$, 
$\eta_\text{read-in} = 39.1 \% \pm 3.5 \%$ and
$\eta_{\text{tot}}= 21.1 \% \pm 1.9 \%$.

\begin{figure}[h]
\begin{center}
\includegraphics[width=.9\columnwidth]{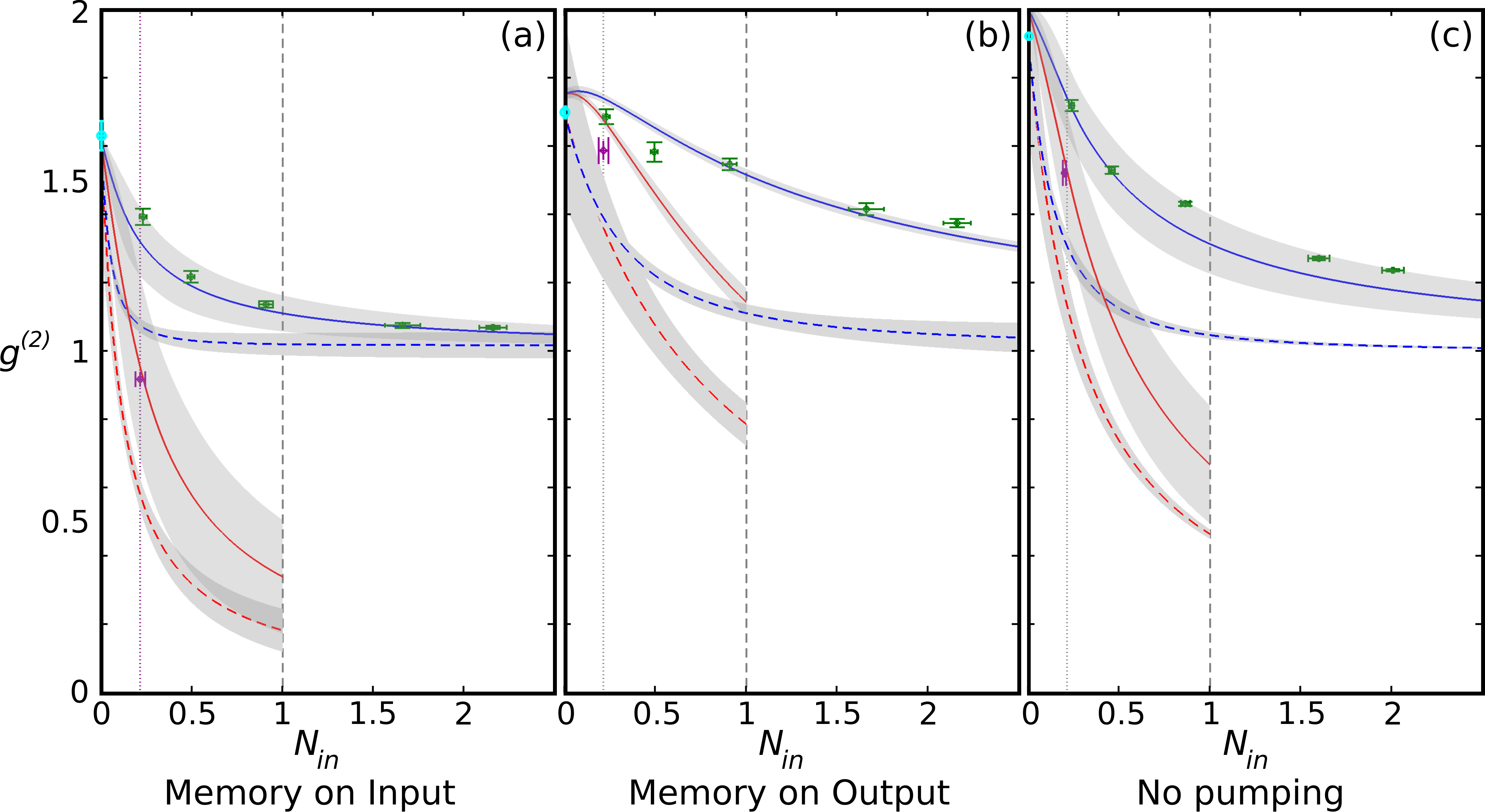}
\caption{{\bf Photon statistics measurements.} 
$g^{(2)}$-values expected from the incoherent interaction model in Eq.~\ref{eq_goldschmidt} (dashed lines) versus the experimental data (points) and the coherent $g^{(2)}$-model prediction (solid lines), for (a) the fields transmitted through the memory at storage, (b) the fields retrieved from the memory at read-out, and (c) the fields transmitted through the memory when the optical pumping beam is blocked. Grey shaded regions denote errors derive from the experimentally measured errors on the variables in Eq.~\ref{eq_goldschmidt} and from propagating uncertainties in the experimental parameters through the FWM model with a Monte-Carlo simulation (see \S \ref{monte}).
The remaining colour coding is as described in Fig.~3 in the main text.
}
\label{fig:supp_g2_plots_vertical}
\end{center}
\end{figure}

\noindent
The results of Eq. \ref{eq_goldschmidt} are shown in Fig. \ref{fig:supp_g2_plots_vertical} alongside the experimental data and the coherent $g^{(2)}$-model predictions. 
Clearly the prediction from the incoherent model significantly underestimates the experimentally measured $g^{(2)}$ data in all three observed cases (memory on input and output time bin as well as optical pumping off).

\noindent
Notably one could argue that the conservation of the photon statistics for the input signal during memory storage is not necessarily a given. 
In such a case, it would thus not be justified to use $g^{(2)}_\text{sig}$ of the input signal in Eq. \ref{eq_goldschmidt} for calculating $g^{(2)}_\text{tot} $ of the memory output, comprising retrieved signal and noise. Instead, a modified $g^{(2)}_\text{sig}$ would need to be used, which is not directly accessible experimentally. 
This potential flaw is circumvented by also measuring $g^{(2)}_\text{tot}$ with the optical pumping switched off (Fig. \ref{fig:supp_g2_plots_vertical} \textbf{c} and Fig.~3 \textbf{c}). 
Here there is no Raman storage, hence no modification of $g^{(2)}_\text{sig}$ should occur. Moreover, $N_\text{sig}$ corresponds directly to the input photon number $N_{\text{in}}$, without any additional memory efficiency factors. 
Since Eq. \ref{eq_goldschmidt} also fails to describe the situation without optical pumping, where all variables are known by direct measurement, we can reject this model and its associated implication. 

\noindent
In conclusion, both light fields cannot be considered as individual entities. 
Consequently, the system can only be correctly described when taking the full dynamics between the Cs atoms and the input light fields into account.

\begin{figure}[h!]
\begin{center}
\includegraphics[width=.8\textwidth]{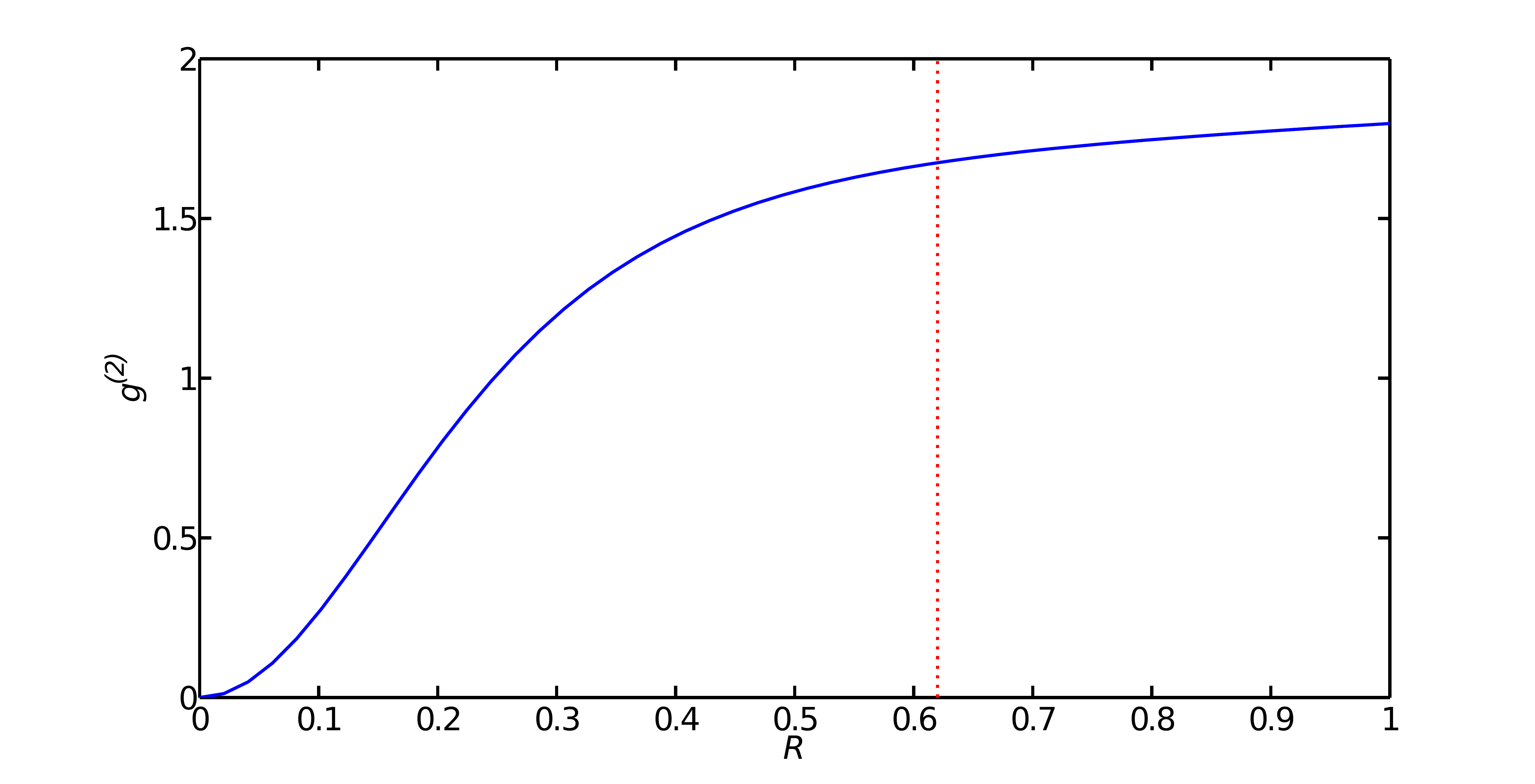}
\caption{Theoretical prediction for the $g^{(2)}$ autocorrelation of the retrieved field after storing a single photon, as the ratio $R = C'/C$ is varied. The vertical dotted line shows the value $R = 0.625$ that describes our experiments.}
\label{fig:r_scan}
\end{center}
\end{figure}

\subsection{Effect of reduced anti-Stokes scattering}

\noindent
The undesired effects of four-wave mixing can be reduced by suppressing the strength of anti-Stokes scattering. To explore this theoretically, we consider the storage and retrieval of single photons with a range of values for the ratio of coupling strengths $R = C'/C$. As shown in Fig.~\ref{fig:r_scan} below, suppressing the anti-Stokes coupling by a factor of 2.5 is sufficient to observe non-classical statistics at the output. If a suppression of 30 could be achieved, the output statistics would be as non-classical as those of the input (i.e. the noise would be dominated by the source, rather than the memory).

\newpage

